
\input amsppt.sty
\magnification=\magstep1
\hsize = 6.5 truein
\vsize = 9 truein

\TagsOnLeft
\NoRunningHeads
\NoBlackBoxes
\TagsAsMath
\catcode`\@=11
\redefine\logo@{}
\catcode`\@=13

\def\label#1{\par%
        \hangafter 1%
        \hangindent .5 in%
        \noindent%
        \hbox to .5 in{#1\hfill}%
        \ignorespaces%
        }

\newskip\sectionskipamount
\sectionskipamount = 24pt plus 8pt minus 8pt
\def\sectionskip{\vskip\sectionskipamount}
\define\sectionbreak{%
        \par  \ifdim\lastskip<\sectionskipamount
        \removelastskip  \penalty-2000  \sectionskip  \fi}
\define\section#1{%
        \sectionbreak   
        \subheading{#1}%
        \bigskip
        }


\define\op#1{\operatorname{\fam=0\tenrm{#1}}} 

        \define         \x              {\times}
        \let            \< = \langle
        \let            \> = \rangle
        \define         \a              {\alpha}
        \redefine       \b              {\beta}
        \redefine       \d              {\delta}
        \redefine       \D              {\Delta}
        \define         \e              {\varepsilon}
        \define         \g              {\gamma}
        \define         \G              {\Gamma}
        \redefine       \l              {\lambda}
        \redefine       \L              {\Lambda}
        \define         \n              {\nabla}
        \redefine       \var            {\varphi}
        \define         \s              {\sigma}
        \redefine       \Sig            {\Sigma}
        \redefine       \t              {\tau}
        \define         \th             {\theta}
        \redefine       \O              {\Omega}
        \redefine       \o              {\omega}
        \define         \z              {\zeta}
        \redefine       \i              {\infty}
        \define         \p              {\partial}

\topmatter
\title Quasiclassical Asymptotics of Solutions to the KZ Equations
\endtitle
\author Nicolai Reshetikhin$^*$ and Alexander Varchenko"
\endauthor
\date February, 1994 \enddate
\thanks The first author was supported by Alfred P.Sloan fellowship and by
NSF grant DMS-9296120.
The second author was supported by NSF Grant DMS-9203929.
\newline $^*$
 Department of Mathematics,  University of California,
        Berkeley, California 94720
\newline {\it E-mail address:} reshetik\@math.berkeley.edu
\newline "  Department of Mathematics, University of North
 Carolina, Chapel Hill, North Carolina 27599
\newline {\it E-mail address:} varchenko\@math.unc.edu
\endthanks
\abstract
The quasiclassical asymptotics of the Knizhnik-Zamolodchikov system is studied.
Solutions to this system in this limit are related naturally to Bethe vectors
in the Gaudin model of spin chains.
\endabstract
\address\hskip-\parindent
        Department of Mathematics \newline University of California \newline
        Berkeley, California 94720\newline and\newline
        Department of Mathematics\newline University of North
        Carolina\newline Chapel Hill, North Carolina 27599
\endaddress
\endtopmatter
\baselineskip=14 pt
\document

\subheading{Introduction}

The Knizhnik-Zamolodchikov differential equation
$$
\frac{\p F}{\p z_i} \ (z) = \frac 1\kappa \ H_i(z) F(z) \ , \qquad
i=1,\dots ,n \ ,
$$
appeared first as a holonomic system of differential equations on
conformal blocks in a WZW model of conformal field theory [KZ]. Here
$F(z_1,\dots ,z_n)$ is a function with values in the tensor product
$V_1\otimes\dots\otimes V_n$ of representations of a simple Lie
algebra $\frak g$, \ $\kappa =k+g$, where $k$ is the central charge of the
model, and $g$ is the dual Coxeter number of the simple Lie algebra
$\frak g$.

One of the remarkable properties of the KZ system is that the coefficient
functions $H_i(z)$ commute and that the form $w = \sum_i H_i(z)dz_i$
is closed:
$$
\frac{\p H_i}{\p z_j} = \frac{\p H_j}{\p z_i} \ , \qquad
[H_i,H_k] = 0 \ .
$$

In this work we study asymptotics of solutions to the
KZ equation when $\kappa$
tends to zero. In this limit, solutions to the KZ equation turn into
normalized eigenvectors of commuting linear operators
$H_1(z),\dots ,H_n(z)$.

There are integral representations for solutions to the KZ equation,
$$
F(z) = \int_{\Cal C} {\op{exp}}(S(t,z)/ \kappa)A(t,z)dt \ ,
$$
where $S(t,z)$ is some multivalued scalar function of $z$ and $t$, \
$A(z,t)$ is a rational
vector valued function, and $\Cal C$ is a cycle
 on which $S(z,t)$ is a
single-valued function of $t$, $t=(t_1,\dots ,t_k)$ [SV]. If $\kappa$
tends to zero, the integral is localized at critical points of $S$
with respect to $t$. It turns out that every nondegenerate  critical point
$t(z)$ gives an eigenvector $A(t(z),z)$ of $H_1(z),\dots, H_n(z)$.

The algebraic Bethe Ansatz is a certain construction of eigenvectors
for a system of commuting operators. The idea of this construction is
to find a vector-valued function of a special
form and determine its arguments in such a way that the value of this
function will be an eigenvector. The equations which determine these
special values of arguments are called the Bethe equations. For more
details about the algelbraic Bethe ansatz see [FT].

One of the systems of commuting operators which can be diagonalized by
 the ABA is the Gaudin model [G] of an inhomogeneous magnetic chain.

It turns out that the functions $A(z,t)$  are exactly those
special functions which appear in the ABA for the Gaudin model and
that the Bethe equations for the Gaudin model coincide with the equations on
critical points of the function $S(z,t)$:
$$
\frac{\p S(z,t)}{\p t_i} = 0 , \qquad  i=1,...,k .
$$
We show that for every nondegenerate critical point $t(z)$:
$$
B(A(t(z),z), A(t(z),z)) = {\text{const}\cdot}
{\text{ Hess}}_t (S(t(z),z))
$$
where $B$ is the Shapovalov form on $V_1\otimes\dots\otimes V_n$ and
Hess is the Hessian of the function $S(t,z)$ as a function of $t$ at
$t=t(z)$. This is an analog of formulae for norms of Bethe vectors in
[K,R].

As an example we consider the case of  $\frak g=sl_2$. We show
that the constructed eigenvectors are paiwise orthogonal with
respect to the Shapovalov form and form a basis in the space of
singular vectors for generic $z$. We describe asymptotics of this basis when
$| z_1 | << |z_2| << ... << | z_n | .$

The first part of the paper contains some general facts about
asymptoticaly flat sections for families of flat holomorphic
bundles. In the second part we recall some basic facts about
Knizhnik-Zamolodchikov equations. In section 3 we recall integral
representations for solutions to the KZ equations from [SV],[V1].
Section 4 contains the description of asymptoticaly flat sections
of the KZ bundles. Section 5 contains a conjecture about the structure
of critical points in the case n=2. In section 6 we analise the
asymptotic behaviour of critical points. Sections 7,8,9
contain detailed analysis of $sl_2$ case. In section 10
we study the Bethe basis for generic configuration of points
$z_1,...,z_n$. Section 11 contains some remarks about the
relation between asymptoticaly flat sections of the KZ bundle
and the Lame functions. In the last section 12 we discuss
branching properties of the Bethe vectors in the Gaudin model.

The main result of this work is Theorem (4.10) and its Corollary (4.16),
it was found in 1990 and
has been the subject of several talks given in 1992-1993.

When this manuscript was completed we learned about the papers [Ba][BF]
where the relation between asymptotic solutions to the KZ system
and the Bethe vectors has been found.

Similarly quasiclassical assymptotics of q-KZ equations [FR] reproduces
Bethe vectors for corresponding spin chains [R2],[TV].

The structure of the Bethe vectors for the Gaudin model is naturally related to
the representation theory of affine Lie algebras at the critical level.
This subject has been studied in [FFR].

One of us (N.R.) is gratefull to B.Feigin , E.Frenkel, I.Frenkel, V.Ginzburg,
and to E.Sklyanin for many interesting discussions.

\bigskip
\subheading{1. Asymptotically flat sections}

Let $\Cal D$ be a ball in $\Bbb C^n$, $\pi : \Bbb C ^N \times
 \Cal D\to\Cal D$ the projection. Let
$$
\nabla_\kappa = \kappa d-\o
$$
be a {\it family of flat holomorphic connections} in $\pi$, \
$\kappa\in\Bbb C$. Here
$$
\o = H_1 dz_1 +\dots + H_n dz_n
$$
where $\{ H_i\}$ are matrix valued functions.

The flatness means that

$$
\kappa d\o + \o\wedge\o = 0  \
$$
for all $\kappa\in\Bbb C$, or
$$
d\o = 0 \ , \qquad \o\wedge\o = 0,
$$
which implies
$$
\frac{\p H_i}{\p z_j} = \frac{\p H_j}{\p z_i} \ , \qquad
[H_i,H_j] = 0    \ ,   \tag 1.1
$$
for all $i$ and $j$.

We are interested in {\it asymptotically flat sections} $F$ for our
family of flat bundles which have the form
$$
F = {\op{exp}}(S/ \kappa)(f_0 + \kappa f_1 +\dots ) \quad
{\text{ for }} \kappa\to 0 \ .    \tag 1.2
$$
Here $S(z_1,\dots ,z_n)$ is a function, $\{ f_j(z_1,\dots ,z_n)\}$ are
sections of $\pi$, and $F$ must be a formal solution to the system of
equations
$$
\nabla_\kappa F = 0  \ .   \tag 1.3
$$

We will call \ exp$(S/ \kappa)f_0$ {\it an asymptotically flat section of
the first order} if there exists a power series (1.2) which provides an
asymptotic solution to (1.3) modulo terms of order $\kappa^2$.

We describe asymptotically flat sections in terms of fundamental
solutions to the system (1.3). First we arrange them into an {\it
asymptotic fundamental solution to} (1.3) which is a matrix with
columns being asymptotically flat sections (1.2) such that its first terms
form a basis of sections of $\pi$.
Assume that
\roster
\item"$\bullet$" linear operators $\{ H_i(z)\}_{i=1}$ are simultaneously
   diagonalizable for each $z\in\Cal D$,
\item"$\bullet$" the spectrum of operators $\{ H_i(z)\}^n_{i=1}$ is simple for
   each $z\in\Cal D$\newline ($\lambda^\a_i \neq \lambda^\beta_i$ for some
   $i=1,\dots ,n$ if $\a \neq \beta$).
\endroster

Let $\phi(z)$ be the matrix with columns being eigenvectors of
$H_i(z)$, \ $i=1,\dots n$, and $\Lambda_i(z)$  the diagonal matrix
with eigenvalues of $H_i(z)$ on the corresponding places in the
diagonal. We have
$$
H_i(z)\phi(z) = \Lambda_i(z)\phi(z)   \tag 1.4
$$
for each $z\in\Cal D$.

\proclaim{(1.5) Proposition } We have the identity
$$
\p_i \Lambda_j - \p_j \Lambda_i = 0   \tag 1.6
$$
\endproclaim

{\smc Proof.} We have the following equalities:
$$
\aligned
\p_i \Lambda_j - \p_j \Lambda_i & = \p_j(\phi^{-1}H_i\phi)
   -\p_i(\phi^{-1} H_j\phi) \\
 & = -\phi^{-1}\p_j \phi\Lambda_i + \Lambda_i \phi^{-1}\p_j\phi
  +\phi^{-1}\p_i \phi\Lambda_j + \Lambda_j \phi^{-1}\p_i\phi\\
 & = -[\phi^{-1}\p_j \phi,\Lambda_i]+[\phi^{-1}\p_i\phi,\Lambda_j] \ .
\endaligned
$$
Clearly the diagonal part of the right-hand side is zero, which
implies  equality (1.6).

\proclaim{(1.7) Corollary } There exists a function $S(z)$
on $\Cal D$ with values in the diagonal $N\times N$ matrices, such that
$$
\Lambda_i (z) = \p_i S(z).   \tag 1.8
$$
\endproclaim

\proclaim{(1.9) Corollary } Let $\{ S_\a (z)\}$ be diagonal elements of
$S(z)$. Then we have
$$
(\phi^{-1}\p_i\phi)_{\a\beta} = N_{\a\beta} ( \p_i S_\a -\p_i S_{\beta})
\ ,   \qquad \a\neq\beta    \tag 1.10
$$
for some matrix $N(z)$ which does not depend on $i$.
\endproclaim

This follows from
 identity (1.6).

The following theorem gives a description of asymptotic fundamental
solutions to system (1.3).

\proclaim{(1.11) Theorem } Let $\Phi(z)$ be the following asymptotic
series expansion:
$$
\Phi(z) = \phi(z) \sum_{K\geq 0}\kappa^K \psi^{(K)}(z)
e^\frac{S(z)}{\kappa} C  \ . \tag 1.12
$$
Then:\newline
(1) $\Phi(z)$ is an asymptotic fundamental solution to the system
$$
\kappa \frac{\p\Phi}{\p z_i} = H_i \Phi \ , \qquad
i=1,\dots N   \tag 1.13
$$
if
\roster
\item"(i)" $C$ is a constant
\item"(ii)" $\phi(z)$ is as in (1.4)
\item"(iii)" functions $\psi^{(K)}(z)$ satisfies the following
   recursive relations:
$$
\psi^{(K)}_{\a\b} = \sum_\gamma N_{\a\gamma} \p_i S_{\a\gamma}
\psi^{(K-1)}_{\gamma\beta}
  +\frac{\p_i \psi^{(K-1)}_{\a\beta}}{\p_i S_\a-\p_i S_\beta} \ ,
\qquad  \a\neq\beta   \tag 1.14
$$
$$
\aligned
& \p_i \psi^{(K)}_{\a\a} +\sum_\gamma N_{\a\gamma}
\psi^{(K)}_{\gamma\a} (\p_i S_\a - \p_i S_\gamma) = 0 \\
& \psi^{(0)}_{\a\b} = \d_{\a\b}
\endaligned    \tag 1.15
$$
   where $N_{\a\b}$ is the same as in (1.10).
\item"(iv)"
$$(\phi^{-1}\p_i\phi)_{\a\a} = 0 \ \ {\text{ for each }}
\a =1,\dots ,N \ . \tag 1.16
$$
\endroster

(2) There exists a solution to (1.14) and (1.15)
which is unique up to
$$
\psi^{(K-1)}_{\gamma\beta} \longrightarrow
 {\text {const}}  \sum_{S+T=K} \psi^{(S)}_{\a\beta} C_\b^{(T)} ,
$$
and
 the matrix $\phi(z)$ has property $(iv)$. Therefore, there exists an
asymptotic
 fundamental solution to system (1.13) .
\endproclaim

{\smc Proof.} (1) We just have to substitute (1.12) into  system (1.13).
Comparing coefficients of asymptotic expansion with respect to
$\kappa$ we obtain the recursive relations for $\psi^{(K)}$:
$$
(\phi^{-1}\p_i\phi)\psi^{(K-1)} + \p_i \psi^{(K-1)} =
[\p_i S,\psi^{(K)}]
$$
and the condition
$$
H_i\phi = \phi \p_i S    \ .
$$

Equations (1.14) and (1.15) are the off-diagonal and diagonal matrix
elements of (1.14), respectively. Let us prove that the system
consisting of (1.14)
and (1.15) has solutions. For this we must show:
$$
(\a) \qquad
\frac{\p_i \psi_{\a\b}^{(K)} +\sum_\g \p_i S_{\a\g}
      N_{\a\g}\psi_{\a\b}^{(K)}}{\p_i S_{\a\b}} =
\frac{\p_j \psi_{\a\b}^{(K)} +\sum_\g \p_j S_{\a\g}
      N_{\a\g}\psi_{\a\b}^{(K)}}{\p_j S_{\a\b}}
$$
$$
(\b) \qquad
\p_j \ \sum_\g N_{\a\g} \psi^{(K)}_{\g\a} \p_i S_{\a\g} =
\p_i \ \sum_\g N_{\a\g} \psi^{(K)}_{\g\a} \p_j S_{\a\g}
$$
where $S_{\a\b} := S_\a - S_\b$.  Let us first prove $(\a)$.

\proclaim{(1.17) Lemma }
$$
\p_i N_{\a\b} = L_{\a\b} \p_i S_{\a\b}   \tag 1.18
$$
for some $L_{\a\b}$.
\endproclaim

{\smc Proof.} Let us recall the definition of $N_{\a\b}$ and compute
$\p_j N_{\a\b}$:
$$
\p_j N_{\a\b} =
\frac{\p_j(\phi^{-1}\p_i \phi)_{\a\b}}{\p_i S_{\a\b}} -
\frac{(\phi^{-1}\p_i\phi)_{\a\b}\p_i\p_jS_{\a\b}}{(\p_i S_{\a\b})^2}
 $$
where $S_{\a\b} := S_\a-S_\b$. Then we have:
$$
\aligned
\p_j(\phi^{-1}\p_i\phi)_{\a\b} & =
    -(\phi^{-1}\p_j\phi \phi^{-1}\p_i\phi)_{\a\b}
    +(\phi^{-1}\p_i\p_j \phi)_{\a\b} \\
& = -\sum_\g N_{\a\g}N_{\g\b} \p_j S_{\a\g} \p_i S_{\g\b}
    +(\phi^{-1}\p_i(\phi[\p_jS,N]))_{\a\b} \\
& = -\sum_\g N_{\a\g}N_{\g\b} \p_j S_{\a\g} \p_i S_{\g\b}
    + ([\p_i S,N][\p_j S,N])_{\a\b}\\
& + [\p_i\p_j S,N]_{\a\b} + [\p_j S,\p_i N]_{\a\b} \ .
\endaligned
$$
Therefore:
$$
\p_j N_{\a\b} =
\frac{\p_j S_{\a\b} \p_i N_{\a\b}}{\p_i S_{\a\b}}
$$
which implies (1.6).

Now we can prove $(\a)$ by induction. Lemma 1.17 provides a base for
induction:
$$
\psi^{(1)}_{\a\b} = N_{\a\b} \ .
$$
Relation (1.18) means that
$$
\p_i \psi^{(1)} = L_{\a\b} \p_i S_{\a\b} .
$$
Assume (1.14) holds for $K$. Let us prove that this implies $(\a)$ and
therefore consistency of (1.14) for $K+1$.
$$
\align
& \p_i S_{\a\b}\left(
  \p_j\psi_{\a\b}^{(K)} +\sum_\g \p_j S_{\a\g}
   N_{\a\g}\psi^{(K)}_{\g\b}\right) - (i \leftrightarrow j) = \\
& = \p_i S_{\a\b} \left( \p_j \left(
  \frac{\sum_\g (\phi^{-1}\p_i\phi)_{\a\g} \psi^{(K-1)}_{\g\b}
      +\p_i \psi^{(K-1)}_{\a\b}}{\p_i S_{\a\b}} \right)
     +\sum_\g \p_j S_{\a\g} N_{\a\g} \psi^{(K)}_{\g\b}\right)
     -(i\leftrightarrow j) \\
& = \left( -\sum_{\g,\t} (\phi^{-1}\p_i\phi)_{\a\g}
    (\phi^{-1}\p_j\phi)_{\g\t} \psi^{(K-1)}_{\t\b}
   + \sum_\g (\phi^{-1}\p_j\p_i\phi)_{\a\g} \psi^{(K-1)}_{\g\b} \right.\\
& \quad + \left. \sum_\g (\phi^{-1}\p_i\phi)_{\a\g} \p_j\psi^{(K-1)}_{\g\b}
  + \p_i\p_j \psi^{(K-1)}_{\a\b} -\psi_{\a\b}^{(K)}
    \p_i\p_j S_{\a\b} +\sum_{\g} \p_i S_{\a\b}
    (\phi^{-1}\p_j\phi)_{\a\g} \psi^{(K)}_{\g\b}\right)
    -(i\leftrightarrow j)\\
& = -\sum_{\g} (\phi^{-1}\p_j\phi)_{\a\g}(\phi^{-1}\p_i\phi)_{\g\t}
   \psi^{(K-1)}_{\t\b}\\
& \quad + \sum_\g (\phi^{-1}\p_i\phi)_{\a\g}\p_j \psi^{(K-1)}_{\g\b}
  + \p_i S_{\a\b} (\phi^{-1}\p_j\phi)_{\a\g} \psi^{(K)}_{\g\b}
  - (i\leftrightarrow j) \\
& = -\sum_\g (\phi^{-1}\p_j\phi)_{\a\g}
   [\sum_\t (\phi^{-1}\p_i\phi)_{\g\t} \psi^{(K-1)}_{\t\b}
   +\p_i \psi^{(K-1)}_{\g\b}] \\
& \quad + \sum_\g (\phi^{-1}\p_i\phi)_{\a\g}
   [\sum_\t (\phi^{-1}\p_j\phi)_{\g\t}\psi^{(K-1)}_{\t\b}
   +\p_i \psi^{(K-1)}_{\g\b}] \\
&  \quad + \p_i S_{\a\b} \sum_\g (\phi^{-1}\p_j\phi)_{\a\g} \psi^{(K)}_{\g\b}
  -\p_j S_{\a\b} \sum_\g (\phi^{-1}\p_i\phi)_{\a\g}\psi^{(K)}_{\a\b}\\
& = -\sum_\g (\phi^{-1}\p_j\phi)_{\a\g} \p_i S_{\a\b}\psi^{(K)}_{\g\b}
    +\sum_\g (\phi^{-1}\p_i\phi)_{\a\g} \p_j
       S_{\g\b}\psi^{(K)}_{\g\b}\\
&\quad +\sum_\g \p_i S_{\a\b} (\phi^{-1}\p_j\phi)_{\a\g}\psi^{(K)}_{\g\b}
  -\sum_\g \p_j S_{\a\b} (\phi^{-1}\p_i\phi)_{\a\g}\psi^{(K)}_{\g\b}\\
& = \sum_\g N_{\a\g} \psi^{(K)}_{\g\b}
  \left( -\p_j S_{\a\g} \p_i S_{\g\b} +\p_i S_{\a\g} \p_j S_{\g\b}
   +\p_i S_{\a\b} \p_j S_{\a\g}-\p_j S_{\a\b} \p_i S_{\a\g}\right) = 0
\endalign
$$
This proves $(\a)$.

Now let us prove $(\b)$:
$$
\aligned
&  \p_j((\phi^{-1}\p_i\phi)\psi^{(K-1)})
 - \p_i((\phi^{-1}\p_j\phi)\psi^{(K-1)}) = \\
& = -\phi^{-1}\p_j\phi \phi^{-1}\p_i\phi \psi^{(K-1)}
    +\phi^{-1}\p_i\phi \p_j \psi^{(K-1)} \\
&\quad +\phi^{-1}\p_i\phi \phi^{-1}\p_j\phi \psi^{(K-1)}
  -\phi^{-1}\p_j\phi \p_i\psi^{(K-1)} \\
& = -[\phi^{-1}\p_j\phi,\phi^{-1}\p_i\phi]\psi^{(K-1)}
    -\phi^{-1}\p_i\phi\cdot \phi^{-1}\p_j \phi\psi^{(K-1)}\\
&\quad  + \phi^{-1}\p_i\phi[\p_jS,\psi^{(K)}]
  + \phi^{-1}\p_j\phi \phi^{-1}\p_i\phi \psi^{(K-1)}
  -\phi^{-1}\p_j\phi [\p_i S,\psi^{(K)}]\\
& = [\p_i S,N][\p_j S,\psi^{(K)}] -[\p_j S,N][\p_i S,\psi^{(K)}]
\endaligned
$$
Diagonal matrix elements of the right-hand side are equal to:
$$
  \sum_\g \p_i S_{\a\b} N_{\a\b} \p_j S_{\b\a}\psi_{\b\a}^{(K)}
- \sum_\b \p_j S_{\a\b} N_{\a\b} \p_i S_{\b\a}\psi_{\b\a}^{(K)} = 0
$$
which proves $(\b)$.
And finally, it is clear that if each $\psi^{(S)}_{\a\b}, \
S=0,\dots  ,K$, satisfies relations (1.14) and (1.15) then
$\psi^{(P)} = \sum^\rho_{S=0} \psi_{\a\b}^{(S)} C_\b^{(P-S)}$
satisfy the same relations. This property reflects the fact that
$\psi^{(K)}(z)$ in (1.12) is determined only modulo multiplication from
the right by a $\kappa$-adic. Theorem 1.11 is proved.

Let $V^*$ be a linear space dual to $V$ and
let $H^*_i: V^*\to V^*$ be
linear operators dual to $H_i$. Denote by $\nabla^*_i$ the
differential operator dual to $\nabla_i$:
$$
\nabla^*_i =-\kappa \ \frac{\p}{\p z_i} - H^*_i   \tag 1.19
$$

Denote by $\< \ \cdot \ , \ \cdot \ \> :V^*\otimes V\to\Bbb C$ the
pairing between $V^*$ and $V$ .

Below we do not assume simplicity of spectrum of $H_i$.

\proclaim{(1.20) Proposition } If $f$ is a solution to (1.3) and $g$ is a
solution to the dual system:
$$
\nabla_i^* g = 0 , \qquad i = 1, \dots , n ,   \tag 1.21
$$
then
$$
\< g,f\> = {\op{const.}}    \tag 1.22
$$
\endproclaim

The proof is obvious.

Consider asymptotic solutions $f$ and $g$ to (1.13) and (1.21),
respectively:
$$
f(z) = {\op{exp}} \left( \frac{S(z)}{\kappa}\right)
    (f_0(z) +\kappa \ f_1(z) +\dots )   \tag 1.23
$$
$$
g(z) = {\op{exp}} \left( \frac{T(z)}{\kappa}\right)
    (g_0(z) +\kappa \ g_1(z) +\dots )   \tag 1.24
$$
Here $S(z)$ and $T(z)$ are two functions which determine
eigenvalues of $H_i$ and $H^*_i$ respectively (see (1.8)).

\proclaim{(1.25) Proposition } (1) If $S - T \neq$ const, we have
$$
\< g_0(z),f_0(z)\> = 0
$$

(2) If $S - T =$ const, we have
$$
\< g_0(z),f_0(z)\> = {\op{const.}}
$$
\endproclaim

{\smc Proof.} Indeed, Proposition 1.20 implies
implies
$$
\< g(z),f(z)\> = {\op{const}}(\kappa) .
$$
On the other hand,
$$
\< g(z),f(z)\> = {\op{exp}}
\left( \frac{S(z)-T(z)}{\kappa} \right)
(\< g_0(z),f_0(z)\> + ...)
$$
This proves (1) and (2).

\bigskip
{\bf (1.26) Remark. } Assume that $V$ has a nondegenerate bilinear form
$(\cdot \ , \ \cdot ): V\otimes V\to\Bbb C$ and the linear operators
$H_i$ are symmetric with respect to the form.
 In this case Proposition 1.20
implies
$$
(f,g) = {\op{const}}(\kappa)   \tag 1.27
$$
if $ f$ is an asymptotic
solution to (1.3) and $g$ is an asymptotic solution to (1.3) in which $\kappa$
is replaced by $-\kappa$.

Proposition 1.25 then
implies that for any two asymptotic solutions $f$ and $g$
to (1.3) having forms (1.23) and (1.24), respectively,
we have the following:
\roster
\item"i)"  \ $(g_0(z),f_0(z)) = 0$ if $S - T \neq $ const,
\item"ii)" \ $(g_0(z),f_0(z)) =$ const   if $S - T = $ const.
\endroster
and, in particular, we have:
$$
(f_0(z),f_0(z)) = {\op{const}}
$$
for every asymptotic solution to $\nabla_\kappa f=0$.

{\bf (1.28) Remark. } Under assumptions of (1.26) one
can easily prove the following generalization
of (1.26). For any two asymptotically flat sections of the first
order $f$ and $g$ we have statements i) and ii) of (1.26).

\bigskip\bigskip

\subheading{2. Knizhnik-Zamolodchikov connections}

Let $V$ be a vector space and let $r_{ij}: \Bbb C ^*  \to {\text{End}}(V)$,
\ $1\leq i < j \leq n$, be functions satisfying the classical
Yang-Baxter equation:
$$
  [r_{ij}(u),r_{ik}(u+v)]+[r_{ij}(u),r_{jk}(v)]
+ [r_{ik}(u+v),r_{jk}(v)] = 0     \tag 2.1
$$
for all triples $i < j < k$ . Assume the following skew-symmetry
condition
$$
r_{ij}(u) = -r_{ji}(-u) \ .   \tag 2.2
$$
Let $d_i : V\to V$, \ $i=s,\dots ,n$,  be linear maps with  the
following property:
$$
[d_i+d_j, r_{ij}(u)] = 0 \ , \qquad [d_i,d_j]=0   \ .
\tag 2.3
$$
Consider
$$
H_i(z_1,\dots ,z_n) = \sum_{j\neq i}
r_{ij}^{V_iV_j} (z_i-z_j)+d_i   \ .
$$
It is easy to verify that this collection of linear operators
 provides a family of flat connections on $\Bbb C^n\backslash\{
{\text{diagonals}}\}$ which we will call the {\it KZ family of flat
connections} corresponding to $\{ r_{ij}(u),d_i\}$.

The {\it KZ equation} is the equation of flat sections:
$$
\nabla _{\kappa} F = 0.
$$

Let $\frak g$ be a complex simple Lie algebra, and
let $(\rho_1,V_1),\dots
,(\rho_n,V_n)$ be $\frak g$-modules, $\rho_i :\frak g \to {\op{End}}(V_i)$.
Define $V=V_1\otimes\dots\otimes V_n$ and
$\Omega\in\frak g\otimes\frak g$ as the element of
$S^2(\frak g) {\overset{\text{inv}}\to\hookrightarrow}
\frak g\otimes\frak g$ corresponding to the Killing form.
If $\{ I_a\}$ is an orthonormal basis in $\frak g$ , then
\newline $\Omega =\sum_{a} I_a\otimes I_a$.

Consider the linear operators $\Omega_{ij}$ acting in $V$ as
$$
\Omega_{ij} = \sum^{{\op{dim }}\frak g}_{a=1}
1\otimes\dots\otimes
\rho_i(I_a) \otimes\dots\otimes \rho_j(I_a)\otimes\dots\otimes 1
$$
for $i < j$ and $\Omega_{ij}=\Omega_{ji}$ for $i > j$.

Define
$$
r_{ij}(u) = \frac{\Omega_{ij}}{u}  \ .
$$
Fix any element $d$ of $\frak g$ and let
$d_i=1\otimes\dots\otimes {\underset{i}\to d}\otimes\dots\otimes
1$.

It is an easy exercise to check that elements $r_{ij}(u)$ and $d_i$ of
End$(V)$ satisfy conditions (2.1)--(2.3) and, therefore, the  linear
operators
$$
H_i = \sum_{j\neq i} \frac{\Omega_{ij}}{z_i-z_j} + d_i
$$
determine a family  of flat connections in the trivial vector bundle
over $\Bbb C^n\backslash\{ {\text{diagonals}}\}$ with fiber $V$.

We consider the case in which all $d_i$ are equal to zero and
$$
H_i = \sum_{j\neq i} \frac{\Omega_{ij}}{z_i-z_j}.   \tag 2.4
$$

Let $e_1,\dots ,e_r, \ f_1,\dots ,f_r, \ h_1,\dots ,h_r$ be the
Chevalley generators of $\frak g$ such that
$$
\aligned
[e_i,f_j] &= \d_{ij} h_i \ , \\
[h_i,e_j] &= a_{ij}e_j   \ , \\
[h_i,f_j] &= -a_{ij}f_j  \ ,
\endaligned
$$
where $(a_{ij})$ is the Cartan matrix.

The map $\tau : \frak g \to \frak g$,
sending $e_1,\dots ,e_r, \ f_1,\dots ,f_r, \
h_1, \dots ,h_r$ to $f_1,\dots ,f_r, \ e_1, \dots ,e_r, \ h_1,$
\newline
$\dots ,
h_r$ , respectively, generate an antiinvolution of $\frak g$.

\proclaim{(2.5) Lemma }  $\tau$ preserves the Killing form on
$\frak g$.
\endproclaim

\proclaim{Corollary}
$$
\tau\Omega =\sum \tau(I_a)\otimes\tau(I_a) = \Omega \ .
$$
\endproclaim

Let $W$ be a $\frak g$ module with highest weight vector $w$.
The {\it Shapovalov
form } $B$ on $W$ is the symmetric bilinear form defined by the
conditions:
$$
B(w,w) =1 , \qquad B(xu,v) = B(u,\tau(x)v) \tag 2.6
$$
for all $u,v\in W$ and $x\in \frak g$.

\proclaim{(2.7) Lemma } Let $W_1$ and $W_2$ be modules
with highest weight, and
let $B_1$ and $B_2$ be their Shapovalov forms. Then
$$
B_1\otimes B_2(\Omega(x\otimes  y),u\otimes v) =
B_1\otimes B_2 (x\otimes y,\Omega(u\otimes v))  \tag 2.8
$$
for all $x\otimes y$, \ $u\otimes v\in W_1\otimes W_2$.
\endproclaim

Assume that $V_1,\dots ,V_n$ are $\frak g$ modules with highest weight. Let
$B_i$ be the Shapovalov form on $V_i$. Set
$$
B = B_i\otimes\dots\otimes B_n \ .   \tag 2.9
$$
$B$ is a symmetric bilinear form on $V$.

\proclaim{(2.10) Lemma } The operators $H_1,\dots ,H_n$ in (2.4)
are symmetric with
respect to $B$.
\endproclaim

If $V_1,\dots ,V_n$ are irreducible finite-dimensional $\frak g$-modules, or
Verma modules with generic highest weights, then the form $B$ is
nondegenerate.

\bigskip\bigskip
\subheading{3. Integral representation for solutions of the KZ
equation}

Let $V_1,..., V_n$ be $\frak g$ modules with highest weight.

Let $\frak h$ be the standard Cartan subalgebra of $\frak g$, \
$\a_1,\dots ,\a_r\in \frak h^*$ the simple roots, $( \ \cdot \ , \ \cdot \
)$ the bilinear form on $\frak h^*$ induced by the Killing form,
$\Lambda_1,\dots ,\Lambda_n$ the highest weights of modules
$V_1,\dots ,V_n$ , respectively.

Set \ $\Lambda=\Lambda_1+\dots +\Lambda_n$.
For $\lambda =(\lambda_1,\dots ,\lambda_r)\in\Bbb Z^r_{\geq 0}$, set
$$
\aligned
& V_\lambda  = \{ v\in V \ |\  h_iv=(\Lambda-\sum^r_{j=1}
\lambda _j\a_j,\a_i)v, \
i=1,\dots ,r \}  \ ,  \\
& {\text{Sing }} V_\lambda = \{ v\in V_\lambda \ |\  e_i v=0, \
 i=1,\dots
,r\} \ .
\endaligned     \tag 3.1
$$

Consider the KZ equation
$$
\kappa  \ dF = \sum_{i < j} \Omega_{ij} \ F \
\frac {d(z_i-z_j)} {(z_i-z_j)} .  \tag 3.2
$$

We will recall a construction of solutions to the KZ equation with
values in a given
\newline Sing $V_\lambda$ [SV].

An arbitrary solution of the KZ
equation is represented as a
 linear combination of the constructed
solutions since the KZ connection commutes
with the $\frak g$ action on $V$ .

Let $k=\lambda_1+\dots + \lambda_n$. Consider the space
$\Bbb C^k$ with
coordinates \ $t_1(1),t_1(2),\dots ,t_1(\lambda_1),$\newline $\dots$ ,
$t_r(1),t_r(2),\dots t_r(\lambda_r)$, the space
\ $\Bbb C^{n+k}$ with coordinates
$t_1(1),t_1(2),\dots ,t_r(\lambda_r)$, \ $z_1,
$ \newline $ \dots ,z_n$, and
the space $\Bbb C^n$ with coordinates $z_1,\dots ,z_n$. Let $p:\Bbb
C^{n+k}\to\Bbb C^n$ be the natural projection. Set
$$
\aligned
\Phi(t,z) & = \prod_{1\leq m < \ell\leq n}
(z_m-z_\ell)^{(\Lambda_m,\Lambda_\ell)/ \kappa}\cdot \\
& \cdot \prod_{m,i,j} (z_m-t_i(j))^{-(\Lambda_m,\a_i)/ \kappa}
  \prod_{\Sb i < \ell{\text{ or}}\\ i=\ell \ {\text {and}} \
j<m \endSb}
(t_i(j)-t_\ell(m))^{(\a_i,\a_\ell)/ \kappa} .
\endaligned     \   \tag 3.3
$$
The function
$\Phi$ is a multi-valued holomorphic function on $\Bbb C^{n+k}$ with
singularities at diagonal hyperplanes.

A {\it monomial of weight} $\lambda$ is an element
$M\in V_\l$ of the form
$$
M=f_{i_1}\dots f_{i_{k_i}} v_1\otimes
 f_{j_1}\dots f_{j_{k_2}} v_2\otimes\dots\otimes
 f_{\ell_1}\dots f_{\ell_{k_n}} v_n \ .   \tag 3.4
$$
Here $v_i$ is the highest weight vector of $V_i$, \ $f$'s are elements
of $\{ f_1,\dots ,f_r\}$.

In [SV], for any monomial $M\in V_\l$, a differential $k$-form
$\eta (M)$ is constructed. The form
$\eta (M)$ is a rational form on $\Bbb
C^{n+k}$ with poles at diagonal hyperplanes.

 Consider the
$V_\l$-valued $k$-form
$$
N=\sum_{M\in V_\l} \Phi \cdot \eta(M)\otimes M   \ . \tag 3.5
$$
In [SV] it is proved that

\roster
\item"(3.6)" For every $i$, the form
$$
\left( \kappa \frac{\p}{\p z_i} -\sum_{j\neq i}
\frac{\Omega_{ij}}{z_i-z_j}\right) N
$$
is the sum of differential of a suitable $(k-1)$-form
and a form which has zero restriction to fibers of the projection $p$.
\item"(3.7)" For every $j$, the form $
e_j N = \sum \Psi\cdot\eta (M)\otimes e_j M $
is the sum of differential of a suitable $(k-1)$-form
and a form which has zero restriction to fibers of $p$.
\item "(3.8)" All forms mentioned in (3.6) and (3.7) have
the shape $\sum \Phi \ \omega (M)\otimes M$
where the sum is over all monomials in $V$,  \ $\{\omega (M)\}$
are suitable rational forms with poles at diagonal hyperplanes.
\endroster

Now let $z=(z_1,\dots ,z_n)$ be a point in $\Bbb C^n$ lying in the
complement to the diagonals. Let $\gamma (z)$ be a $k$-cycle lying in
$p^{-1}(z)\subset \Bbb C^{n+k}$ outside the diagonals. Assume that
$\gamma (z)$ continuously depends on $z$. Then the function
$$
F(z) = \int_{\gamma(z)} N   \tag 3.9
$$
takes values in Sing $V_\lambda$ and satisfies the KZ equation. See a
more precise description of $\gamma (z)$ in [SV, V].

\bigskip\bigskip
\subheading{4. Asymptotic solutions to the KZ equation}

Let $\kappa\to 0$. We use the form $N$ defined in (3.5) and the method
of steepest descent to construct asymptotic solutions to the KZ
equation (3.2).  The function $\Phi$ can be written in the form
$$
\Phi (t,z) = {\op{exp}}(S(t,z)/ \kappa )  \tag 4.1
$$
where
$$
\aligned
S(t,z) & = \sum_{1\leq m < \ell\leq n}
(\Lambda_m,\Lambda_\ell)\ln (z_m-z_\ell)
 \quad - \sum_{m,i,j} (\Lambda_m,\a_i)\ln (z_m-t_i(j))\\
  & \quad +
\sum_{\Sb i < \ell{\text{ or }}\\i=\ell{\text{ and }} j < m\endSb}
  (\a_i,\a_\ell)\ln (t_i(j)-t_\ell(m))
\endaligned       \tag 4.2
$$
Its derivatives have the form
$$
\aligned
 \frac{\p S}{\p z_m} & = \sum_{\ell\neq m} (\Lambda_m,\Lambda_\ell) \
  \frac{1}{z_m-z_\ell} -\sum_{i,j} (\Lambda_m,\a_i)
  \frac{1}{z_m-t_i(j)} \ , \\
 \frac{\p S}{\p t_i(j)} & = \sum_{(\ell,m)\neq (i,j)}
  (\a_i,\a_\ell)\frac{1}{t_i(j)-t_\ell(m)} -\sum_m (\Lambda_m,\a_i) \
  \frac{1}{t_i(j)-z_m} \ , \\
 \frac{\p^2 S}{\p t_\ell(m)t_i(j)} & = (\a_i,\a_\ell) \
  \frac{1}{(t_i(j)-t_\ell (m))^2} \ , \\
 \frac{\p^2 S}{\p t_i(j)^2} & = -\sum_{(\ell,m)\neq (i,j)}
  (\a_i,\a_\ell) \  \frac{1}{(t_i(j)-t_\ell(m))^2} +\sum_m
  (\Lambda_m,\a_i) \ \frac{1}{(t_i(j)-z_m)^2} \ .
\endaligned      \tag 4.3
$$
Set
$$
{\op{Hess}}_t(-S) = {\op{det}}
\left( - \ \frac{\p^2 S}{\p t_i(j)\p t_\ell(m)} \right)\ .  \tag 4.4
$$
For a fixed $z\in\Bbb C^n$, consider the equations of
$t$-critical points
for $S$:

$$
\frac{\p S}{\p t_i(j)} = 0 \ , \qquad
i=1,\dots ,r, \ \ \ j=1,\dots ,\lambda_r \ .   \tag 4.5
$$
Let $t=t(z)$ be a non-degenerate solution to system (4.5)
holomorphically depending on $z$ in a neighborhood of a point
$z^0\in\Bbb C^n$. Set
$$
\overline S(z) = S(t(z),z) \ .   \tag 4.6
$$
Let $B\subset\Bbb C^k$ be a small ball with center at $t(z_0)$.
Let $\d\subset B$ be a $k$-chain such that
$$
{\op{Re }} \ S(t(z_0),z_0)
 > {\op{Re }} \  S(t,z_0)|_{{}_{\p\d}}  \tag 4.7
$$
where $\p\d$ is  boundary of $\d$. This chain is unique in the
sense explained in the remark below. For $z$ close to $z_0$, let
$\d(z)\in p^{-1}(z)$ be the image of the chain $\d$ under the natural
isomorphism $p^{-1}(z)\simeq\Bbb C^k$. Then for $z$ close to $z_0$ we
have
$$
{\op{Re }} \  S(t(z),z) > {\op{Re }} \  S(t,z)|_{{}_{\p\d(z)}}   \tag 4.8
$$
Set
$$
F(z) = \kappa^{-\frac{k}{2}} \int_{\d(z)} N   \tag 4.9
$$
where $N$ is given by (3.5). The function $F$ is defined in a neighborhood of
$z_0$.

\proclaim{(4.10) Theorem }

 1. Let $\kappa \in \Bbb R.$
As $\kappa \to +0$, the function $F$ has an
asymptotic expansion
$$
F(z) = {\op{exp}}(\overline S(z)/ \kappa) \sum^\infty_{m=0}
f_m(z)\kappa^m
$$
where $\{ f_m\}$ are holomorphic functions of $z$.

2. The function $F$ gives an asymptotic solution to the KZ equation.

3. The functions $\{ f_m\}$ take values in ${\op{Sing }} V_\lambda$.
\endproclaim

Part 1 of the theorem is a direct corollary of the method of steepest
descent; see, for example, $\S$11 in [AGV].  Parts 2 and 3 are direct
corollaries of (3.6)--(3.8) and (4.7).

{\bf Remark. } Let
$$
B^- = \{ t\in B \ |\ {\op{Re }}\  S(t(z_0),z_0)
 > {\op{Re }} \  S(t,z_0)\} \ .
    \tag 4.11
$$
It is well known that $H_k(B,B^-,\Bbb Z)=\Bbb Z$ and, moreover, if
$\d$ and $\d'$ are two cycles generating the same element in $H_k$,
then $\int_{\d(z)} N$ and $\int_{\d'(z)}N$ have the same asymptotic
expansions; see $\S$11 in [AGV]. In what follows we assume that $\d$
is a cycle generating $H_k$.

\bigskip
There exist local coordinates $u_1,\dots ,u_k$ in $\Bbb C^k$ centered
at $t(z_0)$ such that
$$
S(u,z_0) = -u^2_1 - \dots -u^2_k + {\text {const}} \ .   \tag 4.12
$$
For such coordinates, $\d$ is a disc
$$
\{ (u_1,\dots ,u_k)\in\Bbb R^k \ |\  u^2_1 +\dots +u^2_k\ \leq \ \e\}
   \tag 4.13
$$
where $\e$ is a small positive number.

\bigskip\bigskip
\subheading{The first term of asymptotics}

Let $F(z)$ be given by (4.9) where $\d$ is defined by (4.13). Then the
first term of asymptotics can be computed explicitly. Namely, let
$M\in V_\lambda$ be a monomial. The form $\eta (M)$ can be written as
$$
\eta (M) = A_M(t,z)dt_1(1)\wedge\dots\wedge dt_r(\lambda_r)
\tag 4.14
$$
where $A_M$ is a rational function. Then the standard formula of the
method of steepest descent states that
$$
f_0(z) = \pm(2\pi)^{k/2} ({\op{Hess}}_t(-S(t(z),z))^{-1/2}
\sum_{M\in V_\lambda} A_M(t(z),z)\ M   \tag 4.15
$$
where the sign depends on  orientation of $\d$.
According to our previous considerations, we have the following
corollary.

\proclaim{(4.16) Corollary }

 The vector $f_0(z)$ lies in
${\op{Sing }} V_\l$. For any $i$, $f_0(z)$ is an eigenvector of $H_i$ with
eigenvalue $\frac{\p S}{\p z_i}(t(z),z)$.
The vector $f_0(z)$ has constant length with respect to the
Shapovalov form $B$ defined in (2.9).
\endproclaim

The last statement can be reformulated as follows. Set
$$
g(t,z) = \sum_{M\in V_\l} A_M(t,z) \ M   \tag 4.17
$$
We will call $g(t(z),z)$ {\it the Bethe vector corresponding to the
critical point} $(t(z),z)$. Then
$$
B( g(t(z),z), g(t(z),z)) = {\op{const}}\cdot{\op{Hess}}_t
(S(t(z),z))  \ .     \tag 4.18
$$

This is a rather surprising relation between critical points of $S$
and the Shapovalov form, cf [K,R].  It is shown in [V] that const
$=1$ for $\frak g=sl_2$.

\bigskip\bigskip
\subheading {5. Conjectures}

Let us study critical points of the function $S$ in the case
$n=2$. Introduce new variables
$$
s_i(\ell) = \frac{t_i(\ell)-z_1}{z_2-z_1} \ .
$$

Obviously,
$$
S(t(s,z),z) \ = \ T(s) \ + \ {\text{const}} \cdot
{\text{ln}}  (z_1 - z_2)
$$
where $S$ is the function defined by (4.2) and
$$
\aligned
T(s) & =  \quad - \sum_{i,j} ( \
(\Lambda_1,\a_i) \ln (-s_i(j)) + (\Lambda_2,\a_i) \ln (1 - s_i(j))\\
  & \quad +
\sum_{\Sb i < \ell{\text{ or }}\\i=\ell{\text{ and }} j < m\endSb}
  (\a_i,\a_\ell)\ln (s_i(j)-s_\ell(m))
\endaligned
$$

Remarkably, equations (4.5) of $t-$critical points become
equations of $s-$critical points for
$\{\{s_i(\ell)\}^{\lambda_i}_{\ell = 1}\}_{i=1}^r$
which do not depend on variables $z_1$ and $z_2$ :
$$
\frac{(\Lambda _1,\a_j)}{s_i(j)} +
\frac{(\Lambda _2,\a_j)}{s_i(j)-1} = \sum_{(\ell,m)\neq(i,j)}
\frac{(\a_i,\a_{\ell})}{s_i(j)-s_\ell(m)} \ .    \tag 5.1
$$

Let $g(t,z)$ be the vector defined by (4.17). Set
$$
G\ = \ g(t,z) / (z_1 - z_2)^k .
$$
Substitute $t = t(s,z)$. Then $G$ becomes a function of $s, z_1, z_2$.

\proclaim{(5.2) Proposition} The vector $G$
does not depend on $z_1,z_2$.
\endproclaim

This proposition follows from the explicit form of functions
$A_M(t,z)$ in (4.17) (see [SV]).

\proclaim { (5.3) Conjecture } For generic weights $\Lambda _1$ and
$\Lambda _2$ the
following is true:
\roster
\item  Each  solution to (5.1) determines a nondegenerate
  critical point of the function $T$.
\item For each solution $s^0$ to (5.1), we have
$$
B(G(s^0),G(s^0)) = {\op{Hess}}_s(T(s^0)) .
$$
\item The product of symmetric groups $\Cal S = \Cal S_{\lambda _1}
 \times  ... \times
\Cal S_{\lambda _r}$ naturally acts on the set of critical points of
the function $T$. We conjecture that for each pair of solutions
$s^0$ and $s^1$ to system (5.1)  which belong to different
$\Cal S-$orbits we have
$$
B(G(s^0),G(s^1)) = 0  .
$$
\item The set of vectors $\{ G(s^0) \}$, where $s^0$ runs through
all $\Cal S-$orbits of critical points of $T$, forms a basis in
Sing$(V_1 \otimes V_2)_{\lambda}$.
\endroster
\endproclaim

Similarly, for dominant integer weights $\Lambda _1$ and
$ \Lambda _2$ we have

\proclaim{(5.4) Conjecture } Let $\tilde V_1$ and $\tilde V_2$
be irreducible highest weight modules with highest weights
$\Lambda _1$ and $\Lambda _2$.
Then (1)--(4) from Conjecture 5.3 hold  in this
case also.
\endproclaim

\bigskip\bigskip
\subheading{6. Asymptotic properties of the critical points}

Let $1 < l < n$ and $v(z)$ be a nondegenerate
critical point with the weight $\sum^l_{i=1} \Lambda_i-\sum^r_{i=1} a_i\a_i$
of the function $S(t,z_1,\dots ,z_l)$.  Here we assume as above that
$z_i\in\Bbb C$, \ $z_i\neq z_j$, and to each point $z_i$ we associate a
highest weight Verma module with highest weight $\L_i$.

Similarly let $u(x)$ be a nondegenerate critical point with
weight $\sum^n_{i=l+1}\L_i-\sum^r_{j=1} b_j\a_j$ of the function
$S(t, x_1,\dots ,x_{n-l-1})$. Here we assume that to each point $x_i$
we associate a highest weighit $\L_{i+l}$.

Let $(\{s_i(j)\}^{c_i}_{j=1})^r_{i=1}$ be a solution to (5.1) with $\L_1$
replaced to $\L_1+\dots +\L_l-\sum^r_{i=1} a_i\a_i $ and $\L_2$ replaced to
$\L_{l+1}+\dots +\L_n-\sum^r_{j=1} b_j\a_j $.

\proclaim{Theorem 6.1} (1) There exists a unique nondegenerate
critical point of $S(t,z_1,\dots ,z_N)$ with weight
$\sum^n_{i=1} \L_i -\sum^r_{j=1} (a_j+b_j+c_j)\a_j$ such that
asymptotically, when $z_{l+1},\dots ,z_n\to\infty$, \
$z_1,\dots ,z_l,\dots ,z_{l+i}-z_{l+1},\dots ,z_M-z_{l+1}$ are finite,
it has the form
$$
\{t(z)\} =
\{v(z_1,\dots z_l)+\Cal O (\frac{1}{z_{l+1}})\} \cup
\{ s_i(j)z_{l+1}+\Cal O (1)\} \cup
\{ u(x_1,\dots ,x_{n-l-1}) + z_{l+1}+\Cal O (\frac{1}{z_{l+1}})\}
$$
where $x_i = z_{l+i+1}-z_{l+1}$.
\endproclaim

{\smc Proof.} We will use the following lemma.

\proclaim{Lemma} Let $t(z) =\sum_{n\geq 0} t^{(n)}z^{-n}$. Then
$$
\frac{1}{t(z)-w} =\frac{1}{t^{(0)}-w}+\sum_{m\geq 1} z^{-m}
\sum_{1\leq n\leq m} \frac{(-1)^n}{(t^{(0)}-w)^{n+1}}
\sum_{\Sb p_1+\dots +p_n=m\\p_i\geq 1\endSb}
t^{(p_1)}\dots t^{(p_n)}    \tag 6.1
$$
\endproclaim

{\smc Proof} is by straightforward computation.

The proof for an arbitrary simple Lie algebra $\frak g$ is absolutely
parallel to the one for $sl_2$. To avoid technicalities we present
here the proof for $sl_2$.

Let $t_\a(z)$ be the asymptotic power series of the following type
when $z_{l+1},\dots ,z_n\to\infty$ and $x_i=z_i-z_{l+1}$ is finite,
$i=l+1,\dots ,n$.
$$
\aligned
t_\a(z) & = v_\a(\tilde z) +\sum_{p\geq 1}
v_\a^{(p)}(\tilde z,x)z^{-p}_{l+1} \ , \qquad \a =1,\dots a \\
t_\a(z) & = z_{K+1}s_{{\a}-a} +\sum_{p\geq 0} s_{{\a}-a}^{(p)}(\tilde z,x)
z^{-p}_{l+1} \ , \qquad \a =a+1,\dots ,a+c \\
t_\a(z) & = z_{l+1} + u_{{\a}-a-c}(x) + \sum_{p\geq 1}
  u_{{\a}-a-c}^{(p)}(\tilde z,x)z^{-p}_{l+1} \ , \qquad
  \a = a+c+1,\dots ,a+c+b
\endaligned      \tag 6.2
$$
where $z=(z_1,\dots ,z_n)$, \ $\tilde z=(z_1,\dots z_l)$, \
$x=(0,x_{l+2},\dots ,x_n)$. Then when $\a =1,\dots ,a$ we have the
following:
$$
\aligned
& \sum^{a+b+c}_{\Sb \b\neq\a\\ \b =1\endSb}
\frac{1}{t_\a(z)-t_\b(z)}  =
\sum^a_{\Sb \b\neq\a\\ \b =1\endSb}
\frac{1}{v_\a(\tilde z)-v_\b(\tilde z)} + \\
& \sum_{M\geq 1} z_{K+1}^{-M}
\left\{ \sum_{1\leq L\leq M} \frac{(-1)^L}{v_{\a\b}^{L+1}}
\sum_{\Sb p_1+\dots +p_L=M\\p_i\geq 1\endSb}
v_{\a\b}^{(p_1)}\dots v_{\a\b}^{(p_L)} - \right. \\
&- \sum_{1\leq L\leq M-1} \frac{(-1)^L}{s_\b^{L+1}}
\sum_{\Sb p_1+\dots +p_L=M-1\\p_i\geq 1\endSb}
s_{\a\b}^{(p_1)}\dots s_{\a\b}^{(p_L)} - \\
& - \left. \sum_{1\leq L\leq M-1} (-1)^L
\sum_{\Sb p_1+\dots +p_L=M-1\\p_i\geq 1\endSb}
u_{\a\b}^{(p_1)}\dots u_{\a\b}^{(p_L)} \right\}
\endaligned         \tag 6.3
$$
Here $v_{\a\b}^{(p)} =v_\a^{(p)}-v_\b^{(p)}$, \
$u_{\a\b}^{(p)} = v_\a^{(p+1)}-u_\b^{(p+1)}$ and
$s_{\a\b}^{(p)}=v_\a^{(p+1)}-s_\b^{(p+1)}$. We have similar
formulas for $\a =a+1,\dots ,a+c$ and for $\a =
a+c+1,\dots ,a+c+b$, and for expansions of
$$
\sum^n_{i=1} \frac{\l_i}{t_\a -z_i}  \ . \tag 6.4
$$
Then one can verify that the equality
$$
\sum^n_{i=1} \frac{\l_i}{t_\a -z_i} =
\sum^{a+b+c}_{\b\neq\a} \frac{2}{t_\a-t_\b}   \tag 6.5
$$
for $t_\a$ being an asymptotic power series (6.2) is equivalent to the
following recursive system of equations:
$$
\aligned
& \sum^a_{\b =1} K_{\a\b} v_\b^{(m)} =
{\text{certain function of }}(v,v^{(1)},\dots ,t^{(m-1)};s,\dots,s^{(m-1)};
u,\dots,u^{(m-1)})\\
&\sum^{a+c}_{\b =c+1} L_{\a\b} s_\b^{(m)} =
{\text{certain function of }}(v,v^{(1)},\dots ,t^{(m-1)};s,\dots,s^{(m-1)};
u,\dots,u^{(m-1)})\\
& \sum^{a+b+c}_{\b =a+c+1} M_{\a\b} u_\b^{(m)} =
{\text{certain function of }}(v,v^{(1)},\dots ,t^{(m-1)};s,\dots,s^{(m-1)};
u,\dots,u^{(m-1)})\\
\endaligned            \tag 6.6
$$
Here
$$
\aligned
& K_{\a\b} = \left\{ \sum^l_{i=1} \frac{\l_i}{(v_\a-z_i)^2}
-\sum_{\b\neq\a} \frac{2}{(v_\a-v_\b)^2}\right\}
\d_{\a\b} - (1-\d_{\a\b}) \frac{2}{(v_\a-v_\b)^2} \\
& L_{\a\b} = \left\{
\frac{\l_1 +\dots +\l_l}{s^2_\a} +
\frac{\l_{l+1}+\dots +\l_l}{(1-s_\a)^2} -\sum_{\b\neq\a}
\frac{2}{(s_\a-s_\b)^2}\right\} \d_{\a\b} -(1-\d_{\a\b})
\frac{2}{(s_\a-s_\b)^2} \\
& M_{\a\b} = \left\{ \sum^n_{i=l+1}
\frac{\l_i}{(u_\a-z_i)^2} -\sum_{\b\neq\a}
\frac{2}{(u_\a-u_\b)^2}\right\} \d_{\a\b}
- (1-\d_{\a\b}) \frac{2}{(u_\a-u_\b)^2}
\endaligned    \tag 6.7
$$
These matrices are nondegenerate according to our assumptions and to
Conjecture 5.3. Therefore, recurrencies determine power series (6.2)
uniquely and these series will satisfy equations (6.5).

This finishes the proof of Theorem 6.1 for $\frak g =sl_2$. For other
simple Lie algebras the proof is absolutely similar.

\bigskip\bigskip

\bigskip\bigskip
\subheading {7. \ $sl_2$-case}

Let $\frak g=sl_2$. In terms of the standard generators
$e,f,h$ we have
$$
\O = \tfrac 12 h\otimes h +e\otimes f +f\otimes e \ .
$$

Let $V_1,\dots ,V_n$ be $sl_2$-modules with highest weights
$\Lambda _1,\dots ,\Lambda _n$ and highest weight vectors $v_1,\dots ,v_n$,
respectively. Set $\Lambda =\Lambda _1+\dots +\Lambda _n$.
For $k\in\Bbb Z_{\geq 0}$,
set
$$
\align
& \ V\ \ = V_1 \otimes \dots \otimes V_n \ , \\
& (V)_k = \{ v\in V\ |\  hv=(\Lambda - k \a, \a )v \} \ , \\
& {\op{Sing}} (V)_k =\{ v \in (V)_k \ |\  ev=0\} \ ,
\endalign
$$
where $\a$ is the simple root.

Consider the space $\Bbb C^{n+k}$ with
coordinates $t_1,\dots ,t_k$, \ $z_1,\dots ,z_k$. The functions $\Phi$
and $S$, defined in (3.3) and (4.2), now have the form

$$
\Phi = \prod_{1\leq m < \ell\leq n}
(z_m-z_\ell)^{(\Lambda _m,\Lambda _\ell)/ \kappa}  \prod_{m,i}
(z_m-t_i)^{-(\Lambda _m,\a)/ \kappa} \prod_{i < j} (t_i-t_j)^{2/ \kappa}
$$

$$
S  = \kappa \ln \Phi  .
    \tag 7.1
$$

 A monomial of weight $k$ is an element $M_K \in (V)_k$ of the form
$$
M_K = f^{k_1}v_1\otimes\dots\otimes f^{k_n}v_n \ ,
 \tag 7.2
$$
$K=(k_1,\dots ,k_n)$, \ $k_1+\dots +k_n=k$.  For a monomial $M_K$, the
differential form $\eta(M_K)$, used in (3.5) and (3.9) to
construct solutions to the KZ equation, has the form
$$
\aligned
& \eta(M_K)  = A_{M_K} dt_1\wedge\dots\wedge dt_k \ , \\
& A_M  = \sum_{\sigma\in S(k,k_1,\dots ,k_n)}\  \prod^k_{i=1} \
\frac{1}{t_i-z_{\sigma(i)}} \ .
\endaligned       \tag 7.3
$$
The sum is over the set $S(k,k_1,\dots ,k_n)$ of maps $\sigma$ from
$\{ 1,\dots ,k\}$ to $\{ 1,\dots ,n\}$ such that for all $m$ the
cardinality of $\sigma^{-1}(m)$ is $k_m$.

Let $t=t(z)$ be a nondegenerate solution to the system of equations
$$
\frac{\p S}{\p t_i} \ (t,z)=0 \ , \qquad
i=1,\dots ,k \ .          \tag 7.4
$$
Then formula (4.15) gives a vector $f_0(z)\in {\op{Sing}}(V)_k$ with
properties indicated in (4.16). The function
exp$(S(t(z),z)/ \kappa)f_0(z)$ is an asymptotically flat section of
first order.

 From now on we assume that $V_1,\dots ,V_n$ are Verma modules. In
$\S$9 we will show that the eigenvectors constructed by (4.15) form a
basis in Sing$(V)_k$ for generic $\Lambda _1,\dots ,\Lambda _n$,
 and   $z_1,\dots
,z_n$. We will show that these vectors are pairwise orthogonal with
respect to the Shapovalov form. $\S$8 contains a preliminary information.

\bigskip\bigskip
\subheading{8. Iterated singular vectors}

Let $W_i$, $i=1,2$, be $sl_2$ modules. Let $\o_i\in W_i$ be a singular
vector of weight $m_i\in\Bbb C$, that is, $e\o_i =0$ and
$h\o_i=m_i\o_i$. For a nonnegative integer $\ell$, the vector
$$
(\o_1,\o_2)_\ell  := \left( \prod^\ell_{j=1}
(m_1+m_2+j+1-2\ell) \right) \cdot
$$
$$
 \cdot \sum^\ell_{p=0} (-1)^p
\left( \matrix \ell \\ p \endmatrix \right)\cdot
 \prod^{p-1}_{j=0} (m_1-j)^{-1}\cdot
       \prod^{\ell-p-1}_{j=0} (m_2-j)^{-1}
f^p\o_1\otimes f^{\ell-p}\o_2
 \tag 8.1
$$
is a singular vector in $W_1\otimes W_2$ of weight $m_1+m_2-2\ell$.
We assume that all denominators in this formula are not zero.

Let $\O$ be the Casimir operator acting on $W_1\otimes W_2$.
$$
(\o_2,\o_2)_\ell {\text{ is an eigenvector of }} \O
{\text{ with eigenvalue }}
 \l(m_1,m_2;\ell) =\tfrac 12 m_1m_2 -\ell(m_1+m_2)+\ell(\ell-1)  .
  \tag 8.2
$$

Now let $V_1,\dots ,V_n$ be $sl_2$ Verma modules with highest weights
$ \Lambda _1,\dots ,  \Lambda_n$
 and highest vectors $v_1,\dots ,v_n$, respectively.
Set $m_i = ( \Lambda _i,\a)$.

For any sequence of non-negative integers $I=(i_2,\dots ,i_n)$, \
$i_2+\dots +i_n=k$, set
$$
v_I = (\dots ((v_1,v_2)_{i_2},v_3)_{i_3},\dots ,v_n)_{i_n}  \tag 8.3
$$
$v_I$ is a singular vector in $V=V_1\otimes\dots\otimes V_n$ of weight
$m_1+\dots + m_n-2k$. We assume that all denominators in $v_I$ are not
zero.  The vector $v_I$ will be called the {\it iterated singular
vector}.
$$
 {\text{For generic }} \Lambda _1,\dots ,\Lambda _n,
 {\text{ the iterated vectors }}
 {\text{form a basis in Sing}}(V)_k
 \tag 8.4 $$

It is easy to write explicitly the nondegeneracy conditions.

For any $\ell =2,\dots ,n$, set
$$
G_\ell =\sum_{i < \ell} \O_{i\ell}      \tag 8.5
$$

\proclaim{(8.6) Lemma} For any $\ell$ and any $I=(i_2,\dots ,i_n)$, the
iterated vector $v_I$ is an eigenvector of $G_\ell$ with eigenvalue
$$
\l(m_1+\dots +m_{\ell-1}-2(i_2+\dots +i_{\ell-1}),m_\ell;i_\ell) .
$$
\endproclaim

The lemma easily follows from the following property of the Casimir
operator:
$$
[\O, 1\otimes x+x\otimes 1] =0
$$
for any $x\in sl_2$.

\proclaim{(8.7) Corollary} For generic $\Lambda _1,\dots ,\Lambda _n$
we have the
following property:

Let $I=(i_2,\dots ,i_n)$, \
$i_2 +\dots +i_n = k$, and $J=(j_2,\dots ,j_n)$, \
$j_2+\dots +j_n = k$, be different sequences. Then there exists
$\ell\in\{ 2,\dots ,n\}$ such that the eigenvalues of vectors $v_I$
and $v_J$ with respect to $G_\ell$ are different.
\endproclaim

\bigskip\bigskip
\subheading{9. Critical points of function $\Phi$}

Let
$$
\Phi_2(t) = \prod^k_{i=1} t_i^{-m_1/ \kappa} (t_i-1)^{-m_2/ \kappa}
\cdot \prod_{1\leq i < j \leq k} (t_i-t_j)^{2/ \kappa}
   \tag 9.1
$$
where $m_1,m_2,\kappa\in\Bbb C$ are parameters. This function is a
special case of the function $\Phi$ defined in (7.1). Let
$S_2 =
\kappa\ln\Phi_2 $.

Let
$$
\l_1 = t_1 + \dots  + t_k , \qquad \l_2 = \sum t_i t_j,
\qquad \dots  , \qquad
 \l_k = t_1 \cdot \dots \cdot t_k
$$
be the standard symmetric functions.
Let $$
\mu_1 = (1-t_1) + \dots  + (1-t_k), \qquad  \mu _2  = \sum  (1-t_i)
(1-t_j) , \qquad \cdots
$$
$$
\mu_k=(1-t_1)\dots (1-t_k) ,
\qquad \d = \prod_{1\leq i < j\leq k} (t_i-t_j)^2 .
$$

\proclaim{(9.2) Theorem, [V], cf. [Sz, 6.7] }

 If
$t=(t_1,\dots ,t_k)$ is a critical point of $\Phi_2$, then
$$
\align
\l_\ell & = \left( \matrix k\\ \ell \endmatrix \right)\cdot
  \prod^\ell_{j=1}
  \frac{(m_1+j-k)}{(m_1+m_2+j+1-2k)}  \ , \\
\mu_\ell & = \left( \matrix k\\ \ell \endmatrix \right)\cdot
  \prod^\ell_{j=1}
  \frac{(m_2+j-k)}{(m_1+m_2+j+1-2k)} \ ,
\endalign
$$
for all $\ell$,
$$
\d = \prod^{k-1}_{j=0} (j+1)^{j+1}
\frac{(-m_1+j)^j (-m_2+j)^j}{(-m_1-m_2+2k-2-j)^{2k-2-j}} ,
$$
$$
{\op{Hess}}(-S_2(t)) = k! \prod^{k-1}_{j=0}
\frac{(-m_1-m_2+2k-2-j)^3}{(-m_1+j)(-m_2+j)} \ .
$$
\endproclaim

Let $t$ be a critical point of $\Phi_2$. Consider the
vector $f_0$ given by Theorem (4.10):
$$
f_0 = \pm(2\pi)^{k/2} ({\op{Hess}}(-S_2(t)))^{-\tfrac 12}
\sum^k_{\ell =0} A_{k-\ell,\ell}
f^{k-\ell} v_1\otimes f^\ell v_2 \ ,   \tag 9.3
$$
$$
A_{k-\ell,\ell} =\sum_{1\leq i_1 < \dots < i_\ell\leq k}
\frac{1}{(t_{i_1}-1)} \dots \frac{1}{(t_{i_\ell}-1)}
\prod_{j\not\in (i_1,\dots ,i_\ell)} \frac{1}{t_j} \ .
$$
Here $v_p$, \ $p=1,2$, is the highest weight vector of the Verma
module $V_p$ with highest weight $m_p\in\Bbb C$.

\proclaim{(9.4) Lemma, [V]} We have
$$
A_{k-\ell,\ell}  = (-1)^\ell \left( \matrix k \\ \ell \endmatrix
\right) \cdot\prod^k_{j=1} (m_1+m_2 +j+1-2k)
 \cdot \prod^{k-\ell -1}_{j=0} (m_1-j)^{-1}
  \cdot \prod^{\ell -1}_{j=0} (m_2-j)^{-1}
$$
for all $\ell$.
\endproclaim

The vector $f_0$ is a singular vector of weight $m_1+m_2-2k$,
 it is proportional to
the vector $(v_1,v_2)_k$ in (8.1).

Now let $\Phi$ be the function defined by (7.1). Set
$m_\ell =(\Lambda_\ell,\a)$ for all $\ell$.

Consider in $\Bbb C^k$ the configuration $\Cal C$ of hyperplanes
$$
\aligned
& H_{ij} :  t_i=t_j \ , \qquad 1\leq i < j \leq k \\
& H_i^\ell :  t_i=z_\ell \ , \qquad i=1,\dots ,k \ ,
                                   \ell =1,\dots ,n \ .
\endaligned      \tag 9.5
$$
Let $T$ be the complement in $\Bbb C^k$ to the union of all
hyperplanes of $\Cal C$. Set
$$
T_{\Bbb R} = T\cap\Bbb R^k \  .
$$

\proclaim{(9.6) Theorem, [V]}
Assume that $z_1,\dots ,z_n$ are real and pairwise different.
Assume that $m_1,\dots ,m_n$ are negative. Then all critical points of
$\Phi$ lie in the union of bounded connected components of $T_{\Bbb
R}$. Each bounded component contains exactly one critical point.
This critical point is nondegenerate.
\endproclaim

Assume that $z_1 < z_2 < \dots < z_n$.

We say that a bounded connected
component is {\it admissible} if it lies in the cone
\newline $ t_1 < t_2 < \dots < t_k$.

Admissible components are enumerated by sequences of nonnegative
integers
 \newline $I=(i_2,\dots ,i_n)$, $i_2+\dots +i_n=k$. The corresponding
component has the form
$$
\Cal D_I(z) = \{ t\in\Bbb R^k|z_1 \! < \! t_1 \! < \dots < t_{i_2}
< z_2 < \dots < z_{n-1} < t_{i_2+\dots +i_{n-1}+1} \! < \! \dots \! <
t_k < z_n\}    \tag 9.7
$$
cf. (8.3). The critical point of $\Phi$ in $\Cal D_I(z)$ is denoted
by $t_I(z)$.

Consider the Bethe vector
$$
g_I(z) = \sum_{k_1+\dots +k_n=k} A_{(k_1,\dots ,k_n)}
(t_I(z),z)f^{k_1}v_1\otimes\dots\otimes f^{k_n}v_n    \tag 9.8
$$
where $A_k(t,z)$ is given by (7.3). $g_I(z)$ is a vector in
Sing$(V)_k$ where $V$ is the tensor product of Verma modules with
highest weights $\Lambda _1,\dots ,\Lambda _n$.

It is well known that for generic $\Lambda _1,\dots ,\Lambda _n$
the dimension of
Sing$(V)_k$ is equal to the number of iterated singular vectors
described in $\S$8 and, therefore, it is equal to the number of
admissible domains.

Consider the set of vectors $\{ g_I(z)\}_{I\in{\op{ Ad }}m}$, where $I$
ranges over the set of sequences of nonnegative
integers $(i_2,\dots ,i_n)$, $i_2+\dots +i_n=k$. In [V] it is proved
that under explicitly written conditions on $\Lambda _1,
\dots ,\Lambda _n$ the
Bethe vectors $\{g_I(z)\}_{I\in{\op{ Ad }}m},$ form a basis in
Sing$(V)_k$. Here we'll describe asymptotics of these vectors for $z_1
<< z_2 \dots << z_n$.

\proclaim{(9.9) Theorem}
Assume that $z_j=s^j$, $j=1,\dots ,n$, and
$s\to +\infty$. Then for any $I=(i_2,\dots ,i_n)$, $i_2+\dots +i_n=k$,
we have
$$
g_I(z(s)) = s^{d(I)}(v_I + O(s^{-1}))
$$
where $v_I$ is given by (8.3) and $d(I)$ is an integer.
Moreover, for any $\ell =2,\dots ,n$, the operator
$H_\ell(z) = \sum_{j\neq\ell} \O_{j\ell}/(z_j-z_\ell)$ has the
following asymptotics:
$$
H_\ell = s^{-\ell}(\O_{1, \ell} +\O_{2, \ell} +\dots +\O_{\ell -1 ,\ell}
+O(s^{-1})) \ .
$$
\endproclaim

\proclaim{(9.10) Corollary}
 The Bethe vectors
$\{ g_I(z)\}_{I\in{\op{ Ad }}m}$ form a basis in ${\op{Sing}}(V)_k$
for generic $z_1,\dots ,z_n$ and generic $\Lambda _1,\dots ,
\Lambda _n$.
\endproclaim

\proclaim{(9.11) Corollary}
 Let $\l^\ell_I(z)$ be the eigenvalue of
$g_I(z)$ with respect to $H_\ell(z)$, then
$$
\g^\ell_I(z(s)) = \l(m_1+\dots +m_{\ell -1}-2
(i_2+\dots +i_{\ell -1}),m_\ell;i_\ell) + O(s^{-1})
$$
for $\ell > 1$, see (8.6).
\endproclaim

\proclaim{(9.12) Corollary}
 For generic $\Lambda _1,\dots ,
\Lambda _n$ we have the
following property:

Let $I$ and $J$ be different admissible sequences.
Then there exists $\ell$ such that the eigenvalues of the Bethe
vectors $g_I(z(s))$ and $g_J(z(s))$ with respect to $H_\ell(z)$ are
different for $s >> 1$.
\endproclaim

\proclaim{(9.13) Corollary}
  Let $I$ and $J$ be different admissible
sequences. Then the Bethe vectors  $g_I(z(s))$ and $g_J(z(s))$ are
orthogonal with respect to the Shapabalov form $B$.
\endproclaim

{\bf (9.14) Remark.}  It is shown in [V] that
$$
B(g_I(z(s)),g_I(z(s))) = {\op{Hess}}_t(S(t_I(z),z)) \ .
$$

{\smc Proof of the theorem.} The statement
on $H_\ell$ is trivial. We prove the
statement on $g_I$. Make a change of variables: $t_j =s^\ell u_j$ if
$i_1+\dots +i_{\ell -1} < j \leq i_2 +\dots +i_\ell , \ell = 2, ... ,
n$ (
 we assume that $i_1 = 0$ ) . For any $\ell
=2,\dots ,n$, introduce a function
$$
S_\ell(u) = -\sum^{i_\ell}_{j = i_{\ell - 1} + 1}
(a_\ell \ln u_j +m_\ell \ln(u_j-1) ) + 2
\sum_{i_{\ell - 1} + 1 \leq i < j \leq i_\ell} \ln (u_i - u_j)
$$
where $a_\ell = m_1 +\dots + m_{\ell -1}-2(i_2+\dots + i_{\ell -1})$.
Then we have
$$
S(t(u)) = A \ln s + S_2(u) +\dots + S_n(u) +O(s^{-1})  \tag 9.15
$$
for some number $A$. This formula and the explicit formula for $g_I$
imply the theorem.

Assume that $m_\ell =(\Lambda _\ell,\a)$, $\ell =1,\dots ,n$, are positive
integers. Let $V_1,\dots ,V_n$ be irreducible $sl_2$ modules with
highest weights $\Lambda _1,\dots ,\Lambda _n$, respectively.

Apply the construction of Section 4 to this situation. For any $t-$critical
point $t=t(z)$ of the function $\Phi$ we get a Bethe vector
$g(t(z),z)\in {\op{Sing}}(V)_k$  given by (9.8).

  We say that a
$t-$critical point $t=t(z)$ is
{\it nontrivial} if $g(t(z),z)\neq 0$ , otherwise we will call
 it {\it trivial}.

The set of
$t-$critical points of $\Phi$ is invariant with respect to the group of
permutations of coordinates $t_1,\dots ,t_k$.  Critical points lying
in the same orbit give the same eigenvector.

 Consider the family
$\{ g(t(z),z)\}$ of the Bethe vectors where $t(z)$ runs through the
set of orbits of critical points.

\proclaim{(9.16) Theorem}
For any $z$ every trivial $t$-critical point is
degenerate  and all Bethe vectors $\{ g(t(z),z)\}$ are pairwise orthogonal
with respect to the Shapovalov form. For generic $z$, all nontrivial
critical points are nondegenerate and the corresponding Bethe vectors
form a basis in ${\op{Sing}}(V)_k$.
\endproclaim

{\smc Proof.} Trivial critical points are degenerate by (9.14).
All Bethe vectors are pairwise orthogonal by (9.13). Therefore, it
suffices to show that there are at least dim Sing$(V)_k$ orbits of
nondegenerate critical points for generic $z$.

\bigskip
Let $I=(i_2,\dots ,i_n)$ be a sequence of nonnegative integers such
that $i_1+\dots +i_n =k$. We say that $I$ is a
{\it good sequence} if for
any $\ell =2,\dots ,n$ we have
$$
i_{\ell} \leq \min(m_1+\dots +m_{\ell -1}-2(i_2+\dots +i_{\ell
-1}),m_\ell) \ .
$$
The number $N$ of good sequences is equal to  dimension of
Sing$(V)_k$.

We show that there are at least $N$ orbits of nondegenerate critical
points if $z_\ell =s^\ell$, $\ell =1,\dots ,n$, and $s$ tends to
$+\infty$.

 Let $I$ be a good sequence. Make a change of variables
$t_j =s^\ell u_j$ if $i_1+\dots +i_{\ell -1} < j \leq i_2+\dots +i_\ell$.
Then we have formula (9.15).

 Critical points of $\Phi$ coincide with
critical points of $S(t(u))$. The critical point equations for $S(t(u))$
are a deformation of the critical point equations for $S_2(u)+\dots
+S_n(u)$. Critical points of $S_2+\dots +S_n$ are described by (9.2).

Let $u_I$ be a critical point of $S_2+\dots +S_n$. Let $u_I(s)$ be the
critical point of $S(t(u))$ which is the deformation of $u_I$. Let
$t_I(s)$ be the corresponding critical point of $\Phi$ and
$g_I(t_I(s), z(s))$ the corresponding Bethe vector. Then
$$
g(t_I(s),z(s)) = s^{d(I)} (v_I +O(s^{-1}))
$$
where $d(I)$ is some integer. This describes asymptotics of
$\{ g(t(z),z)\}$ and proves the theorem.

\bigskip\bigskip
\subheading{10. Trees, asymptotic zones, and Bethe bases}

Let $V_1,\dots ,V_n$ be finite-dimensional irreducible $sl_2$ modules.
Consider the family of Bethe vectors $\{ g(t(z),z)\}\in
{\op{Sing}}(V)_k$. We proved that for
$$
|z_2-z_1| << |z_3-z_2| << \dots << |z_n-z_{n-1}|   \tag 10.1
$$
the Bethe vectors form a basis in Sing$(V)_k$. Moreover, the Bethe
vectors can be enumerated by good sequences $\{ I\}$ in such a way
that for every $I$,
$$
g(t_I(z),z) \sim {\op{const}}(z) \cdot v_I    \tag 10.2
$$
where $v_I$ is the corresponding iterated vector, const$(z)$ is a
scalar function, const $\neq 0$.

These facts have the following generalization. {\it An $n$-tree} is a
planar tree with $n$ tops, one root, and $n-1$ internal triple
vertices.  See an example in the figure below.
\newline \
\newline \
\newline \
\newline \
\newline \
\newline \

\vskip 1.25 truein

\centerline{Figure 1. An $n$-tree $T_0$.}
\bigskip

\noindent

The $n$-tree $T_0$, shown in the figure, defines an
``asymptotic zone'' described  by (10.1) and a basis
$\{ v_I\}\in{\op{ Sing}}(V)_k$ where $I$ runs through the good sequences.

Similarly, every $n$-tree $T$ defines two objects: an asymptotic zone,
described by
\newline inequalities similar to (10.1), and a distinguished basis
$\Cal B_T$ in Sing$(V)_k$ consisting of iterated vectors where
iterations are defined according to the shape of the tree.

\proclaim{(10.3) Theorem}

For any $n$-tree $T$, we restrict the Bethe
vectors
$\{ g(t(z),z)\}$ to the asymptotic zone defined by $T$. Then the Bethe
vectors can be enumerated by elements of the basis $\Cal B_T$ in such
a way that for every $v\in\Cal B_T$ we have
$$
g_v(t(z),z) \sim {\op{const}}(z) \cdot v
$$
where ${\op{const}}(z)$ is a scalar function, ${\op{const}} \neq 0$.
\endproclaim

The proof is the same as for (9.16).

\bigskip
Now consider an arbitrary complex simple Lie algebra $\frak g$ instead of
$sl_2$. Assume that Conjecture (5.4) is true for $\frak g$.

Let $V_1,\dots ,V_n$ be finite-dimensional irreducible $\frak g$ modules.
Consider the KZ equation with values in Sing $(V)_\l$, defined by
(3.2). Let $\{g(t(z),z)\}\in$ Sing$(V_\l)$ be the Bethe vectors
constructed in Section 4.

For any $n$-tree $T$ we can construct a distinguished basis $\Cal B_T$
in Sing$(V)_\l$. The basis consists of the iterated vectors where
iterations are constructed according to the shape of the tree, and at
each step of the iteration we use the eigenvectors discussed in
Conjecture (5.4). For example, for the tree shown in the figure and
$g=sl_2$, the iteration procedure is given by (8.3) and the
eigenvector used in each step of iteration (8.3) is given by (8.1).

\proclaim{(10.4) Theorem} Assume that Conjecture (5.4) is true for
$\frak g$.
For any $n$-tree $T$, we restrict the Bethe vectors
$\{ g(t(z),z)\}$ to the asymptotic zone defined by $T$. Then for every
$v\in\Cal B_T$ there exists a Bethe vector $g(t(z),z)$ such that
$g(t(z),z)\sim{\op{ const}}(z)
 \cdot v$ where ${\op{const}}(z)$ is a scalar
function, ${\op{const}} \neq 0$.
\endproclaim

The proof is the same as for (9.16).

\bigskip\bigskip
\subheading{11. Bethe vectors and Lame functions}

Consider the following problem [Sz, 6.8].

\proclaim {(11.1) Problem} Let $A(t)$ and $B(t)$ be given polynomials of
degree $n$ and $n-1$, respectively.  To determine a polynomial $C(t)$
of degree $n-2$ such that the differential equation
$$
A(t)y''(t) -  B(t)y'(t)  +  C(t)y(t) = 0   \tag 11.2
$$
has a solution which is a polynomial of preassigned degree $k$.
\endproclaim

A polynomial solution $y(t)$ is called a {\it Lame function}.
Hein (1878) proved that, in general, there are exactly
$\left(\matrix k+n-2\\ k\endmatrix\right)$ determinations of $C(z)$.

{\bf Example}. Let $A=(1-t^2)$, \ $B=(\a-\b +(\a+\b+2)t)$, then
$C=k(k+\a+\b+1)$ and the corresponding polynomial solution of degree
$k$, normalized by the condition
$y(1) = \left( \matrix k+\a \\ k\endmatrix\right)$, is called {\it the
Jacobi polynomial} and is denoted by $P_k^{(\a,\b)}(t)$.

Let $A=(t-z_1)\dots (t-z_n)$,
$$
\frac{B(t)}{A(t)} = \frac{m_1}{t-z_1} +\dots +
\frac{m_n}{t-z_n}    \tag 11.3
$$

\proclaim{(11.4) Theorem, Stieltjes [Sz, $\S$6.8]}
 Let $A$ and $B$ be
given polynomials of degree $n$ and $n-1$, respectively.
Then there exists a polynomial $C$ of degree $n-2$ and a polynomial
solution $y=(t-t_1)\dots (t-t_k)$ of (11.2) if and only if
$t=(t_1,\dots ,t_k)$ is a critical point of the function
$$
\Psi(t_1,\dots ,t_k) = \prod^k_{j=1} \ \prod^n_{i=1}
(t_j-z_i)^{-m_i/ \kappa} \prod_{1\leq i < j \leq k} (t_i-t_j)^{2/
\kappa}
$$
where $\kappa$ is an arbitrary nonzero number.
\endproclaim

Therefore, having a Lame function $y=(t-t_1)\dots (t-t_k)$ we  can
define a Bethe vector $g(z)\in {\op{Sing}}(V)_k$ by formulae (9.8)
and (7.3). Here $V=V_1\otimes\dots\otimes V_n$, \ $V_j$ is an $sl_2$
highest weight module with highest weight $m_j$.

It turns out that if $V_j$, $j=1,\dots ,n$,  is the Verma module with
highest weight $m_j$, then, in general, a Bethe vector defines a Lame
function, and, therefore, Problem 11.1 is equivalent to the problem of
diagonalization in Sing$(V)_k$ of the KZ operators
$$
H_i(z) = \sum_{j\neq i} \frac{\O_{ij}}{z_i-z_j} \ , \qquad
i=1,\dots ,n \ .
$$

Namely, let $g(z)$ be a Bethe vector given by (9.8) and (7.3). Then
the vector
$$
\prod^k_{j=1} \ \prod^n_{i=1} (t_j-z_i) g(z) =
\sum_{k_1+\dots +k_n=k} B_{k_1,\dots ,k_n}
f^{k_1}v_1\otimes\dots\otimes f^{k_n}v_n
$$
polynomially depends on $t_1,\dots ,t_k$, \ $z_1,\dots ,z_n$.

For a sequence $L=(\ell_1\geq\dots\geq\ell_k)$ of positive integers
define a symmetric polynomial $p_L$ by
$$
p_L(t_1,\dots ,t_k) = \sum t^\a
$$
where the sum is over all permutations $\a$ of the sequence
$(\ell_1-1,\dots ,\ell_k-1)$. We have
$$
B_K(z,t) =\sum_L M_{K,L}(z) \cdot p_L(t) \ .
$$
Here $K=(k_1,\dots ,k_n)$, \ $k_1+\dots +k_n=k$, \ $k_j\in\Bbb Z_{\geq
0}$. The sum is over $L=(\ell_1\geq\dots\geq\ell_k)$, \ $n\geq\ell_1$.
Denote by $\Cal K_{k,n}$ (respectively $\Cal L_{k,n}$) the set of such
$K$ (respectively $L$).

For every $K,L$, the coefficient $M_{K,L}$ is a polynomial in $z$
depending only on $k,n$ and independent of $m_1,\dots ,m_n$. It is
easy to see that $\#\Cal K_{k,n}=\#\Cal L_{k,n}$.

\proclaim{(11.5) Theorem}
We have
$$
{\op{det}}(M_{K,L}(z))\not\equiv 0 .
$$
\endproclaim

\proclaim{Corollary} If ${\op{det}}(M_{K,L}(z))\neq 0$ for some $z$,
then the coefficients $B_K(t,z)$ of a Bethe vector uniquely determine
the symmetric functions $\{ p_L(t)\}$, the orbit of the critical point
$t=(t_1,\dots ,t_k)$, and the corresponding Lame function, cf. [S].
\endproclaim

{\bf Remark.} In several examples, we have
$$
{\op{det}}(M_{K,L}(z)) ={\op{const}} \prod(z_i-z_j)^a
$$
for suitable $a$.

\bigskip
{\smc Proof.} Let $z_j=s^j$, \ $j=1,\dots ,n$. Then
$(M_{K,L}(z(s)))$ is a polynomial in $s$. Let
$d_{K,L}={\op{deg}}_s M_{K,L}(z(s))$, \ $d_L=\max_K d_{K,L}$. Let
$m_{K,L}$ be the coefficient of $s^{d_L}$ in $M_{K,L}$. We'll show
that det$(m_{K,L})\neq 0$. This would imply the theorem.

Define the lexicographical order in $\Cal L_{k,n}$ by the rule:
$(\ell_1\geq\dots\geq\ell_k) > (\ell'_1\geq\dots\geq\ell'_k)$ if
$\ell_k=\ell'_k,\dots ,\ell_{j+1}=\ell'_{j+1}$, \ $\ell_j > \ell'_j$
for some $j=1,\dots ,k$.

Define a bijection $\Cal L_{k,n}\to \Cal K_{k,n}$ by the rule
$(\ell_1\geq\dots\geq\ell_k) \mapsto\sum^k_{j=1} (0,\dots
,0,1_{\ell_j},0,\dots ,0)$. This bijection and the lexicographical
order on $\Cal L_{k,n}$ induce an order on $\Cal K_{k,n}$. It is easy
to see that the matrix $(m_{K,L})$ is a triangular matrix with nonzero
diagonal elements and, therefore, det$(m_{K,L})\neq 0$.

An explanation of a connection between the Bethe vectors and the Lame
functions is given in the Sklyanin paper [S] in which the
tensor product $V$ of $sl_2$ modules is realized in a suitable space
of jets of functions of one variable, the $sl_2$ action is realized by
differential operators, and the KZ operators are realized in terms of
differential operators (11.2).

\bigskip\bigskip
\subheading{12. Branching of the Bethe vectors, Jordan blocks of the
KZ operators}

The Bethe vectors discussed in this work may have branching outside
diagonals $z_i=z_j$. Here is an example. Let $V$ be an $Sl_2$ module
with highest weight $m\neq 0$. Consider the Bethe equation
corresponding to $(V\otimes V\otimes V)_1$:
$$
\frac{1}{t-z_1} + \frac{1}{t-z_2} + \frac{1}{t-z_3} =0 \ .
$$
For $(z_1,z_2,z_3)=(0,1,s)$, this equation gives
$3t^2-2(1+s)+s=0$. Its discriminant has two roots $s_1$ and $s_2$.
Therefore, for $s\neq s_1,s_2$ there are two Bethe vectors and they
form a basis in the two-dimensional Sing$(V\otimes V\otimes V)_1$.
When $s$ goes around $s_j$, $j=1,2$, the Bethe vectors interchange
positions. For $s=s_j$, the KZ operators $H_\ell(z(s_j))$, $\ell
=1,2,3$, have a two-dimensional Jordan block. Their single eigenvector
is the Bethe vector corresponding to the single critical point.

\bigskip\bigskip
\centerline{\bf References}
\bigskip

\label{[AGV]} V. I. Arnold, S. M. Gusein-Zade, and A. M. Varchenko,
{\it Singularities of Differentiable Maps}, volume II, Birkhauser,
1988.
\label{[B]} H. Bethe, {\it Z. Phys.} {\bf 71} (1931), 205.
\label{[Ba]} H.M.Babujian, Off-shell Bathe ansatz equation and N-point
correlators in the SU(2) WZNW theory, preprint Bonn-HE-93-22.
\label{[BF]} H.M.Babujian, R.Flume, Off shell Bethe ansatz equation
for Gaudin magnets and solutions of Knizhnik-Zamolodchikov
equations, preprint Bonn-HE-93-30.
\label{[FFR]} Gaudin model, Bethe ansatz and correlation functions at the
critical level, preprint (1994), 39 pages.
\label{[FR]} I.Frenkel, N.Reshetikhin, Quantum affine algebras and
holonomic difference equations, Comm. Math. Phys. 146 (1992), 1-60.
\label{[G]} M. Gaudin, ``Diagonalizations d'une classe
d'hamiltoniens de spin,'' {\it Jour. de Physique} {\bf 37}, no. 10
(1976), 1087--1098.
\label{[FT]} L. Faddeev and L. Takhtajan, {\it Usp. Math. Nauk.} {\bf
34}, N5 (1979), 13--63 (in Russian).
\label{[K]} V. Korepin, ``Calculation of norms of Bethe wave
functions,'' {\it Comm. Math. Phys.} {\bf 86} (1982), 391--418.
\label{[KZ]} V. Knizhnik and A. Zamolodchikov, ``Current algebra and
Wess-Zumino models in two dimenisons,'' {\it Nucl. Phys.} {\bf B 247}
(1984), 83--103.
\label{[R1]} N. Reshetikhin, ``Calculation of the norms of Bethe
vectors in models with SU(3)-symmetry,'' {\it Zapiski Nauch. Semin.
LOMI} {\bf 150} (1986), 196--213.
\label{[R2]} N.Reshetikhin, Jackson type integrals, Bethe vectors, and
 solutions to a difference analog of the Knizhnik-Zamolodchikov system,
Lett. Math. Phys. 26 (1992), 153-165.
\label{[S]} E. Skylanin, ``Separation of variables in the Gaudin
model,'' {\it Jour. of Sov. Math.} {\bf 47}, 2473--2488.
\label{[Sz]} G.Szego, Orthogonal Polynomials, AMS, 1939.
\label{[TV]} V. Tarasov and A. Varchenko, ``Jackson integral
representations for solutions of the KZ equation,'' preprint:
RIMS--949 (1993).
\label{[SV]} V. Schechtman and A. Varchenko, ``Arrangements of
hyperplanes and Lie algebra homology,'' {\it Invent.  Math.} {\bf 106}
(1991), 139--194.
\label{[V1]} A. Varchenko, ``Multidimensional hypergeometric functions
and representation theory of Lie algebras and quantum groups,''
preprint (1992), 401 pages.
\label{[V2]} A. Varchenko, ``Critical points of the product of powers
of linear functions and families of bases of singular vectors,''
preprint (1993), 15 pages.

\enddocument